# Light matter interaction in transition metal dichalcogenides and their heterostructures

Ursula Wurstbauer, Bastian Miller, Eric Parzinger, and Alexander W. Holleitner

*Walter Schottky Institut und Physik Department, Am Coulombwall 4a, Technische Universität München, D-85748 Garching, Germany*

*Nanosystems Initiative Munich (NIM), Schellingstr. 4, 80799 München, Germany*

*Abstract:*

The investigation of two-dimensional van der Waals materials is a vibrant, fast moving and still growing interdisciplinary area of research. Two-dimensional van der Waals materials are truly two-dimensional crystals with strong covalent in-plane bonds and weak van der Waals interaction between the layers with a variety of different electronic, optical and mechanical properties. A very prominent class of two-dimensional materials are transition metal dichalcogenides and amongst them particularly the semiconducting subclass. Their properties include bandgaps in the near-infrared to the visible range, decent charge carrier mobility together with high (photo-)catalytic and mechanical stability and exotic many body phenomena. These characteristics make the materials highly attractive for both fundamental research as well as innovative device applications. Furthermore, the materials exhibit a strong light matter interaction providing a high sun light absorbance of up to 15% in the monolayer limit, strong scattering cross section in Raman experiments and access to excitonic phenomena in van der Waals heterostructures. This review focuses on the light matter interaction in $MoS_2$, $WS_2$, $MoSe_2$, and $WSe_2$ that is dictated by the materials complex dielectric functions and on the multiplicity of studying the first order phonon modes by Raman spectroscopy to gain access to several material properties such as doping, strain, defects and temperature. Two-dimensional materials provide an interesting platform to stack them into van der Waals heterostructures without the limitation of lattice mismatch resulting in novel devices for application but also to study exotic many body interaction phenomena such as interlayer excitons. Future perspectives of semiconducting transition metal dichalcogenides and their heterostructures for applications in optoelectronic devices will be examined and routes to study emergent fundamental problems and many-body quantum phenomena under excitations with photons will be discussed.

## 1. General introduction to 2D materials and in particular semiconducting transition metal dichalcogenides (SC-TMDs)

In two-dimensional charge carrier systems electrons can freely move within the 2D plane, however, their motion is restricted in the third dimension by quantum mechanics. Besides the constant density of states in (quasi) two-dimensional systems, high charge carrier mobility and reduced screening are both of technological and fundamental physical relevance. The reduced phase space causes also enhanced quantum effects and is therefore of importance for many body phenomena such as Coulomb driven formation of excitons - bosonic quasiparticles of a bound electron-hole pair. A prominent example for such systems are highly mobile two-dimensional electron systems confined in silicon inversion layers or GaAs quantum wells, where increased electron correlations accounting for the observation of fractional Quantum Hall effect, electron solids and other correlated phases at low temperatures and high magnetic fields [1]. The dimensionality-related phenomena are even more striking in real two dimensional (2D) materials that have not only 2D character in momentum space, but also in real space. The three dimensional parental of 2D materials are van der Waals (a.k.a. layered) materials characterized by strong covalent or ionic in-plane bonds and weak van der Waals coupling between individual layers that scales with the distance according to $1/d^6$. As a result, single or few layers can be peeled off from bulk crystals with an adhesive tape as demonstrated for the first time for graphene in 2004 [2] and also for other 2D materials [3].

Graphene's peculiar linear band-structure resulting in relativistic Dirac fermions with record room temperature mobility in combination with a series of further superior properties such as e.g. record breaking mechanical strength, stiffness, flexibility, heat conductance, broadband light absorption and chemical inertness [4] together with its ease of access by only a piece of graphite and an adhesive tape has engaged the avalanche-like increase in research efforts of laboratories all over the globe. Despite its superior properties, the lack of a band gap limits graphene's suitability in the areas of digital electronics with required high on-off ratios for opto-/electronic and light emitting devices as well as for solar energy conversion applications [5–7]. Inspired by graphene, manifold other 2D materials have been explored and the list of theoretically predicted and experimentally realized atomistic thin crystals is steadily growing scoring several hundreds of different derivatives that can be classified in 'families' of 2D materials with different physical properties even varying with the number of layers [8–10]. The main families are the graphene family with graphene and their chemical modifications as well as hexagonal boron nitride (hBN), the layered chalcogenides and the layered oxides [11]. 2D materials span the electronic classes from metals, half-metals, insulators, semiconductors, superconductors, to magnetic and ferroelectric materials and even more exotic ones such as topological insulators [8–13]. Consequently, for almost any application a conceptually suitable 2D material is likely to exist.

A major advance towards the realization of future 2D materials based flexible nano- and optoelectronic devices was the observation by Mak *et al*. [14] in 2010. A single layer of a $MoS_2$ crystal turns into a direct gap semiconductor emitting in the visible range [14,15], followed by the realization of a single layer $MoS_2$ field effect transistor (FET) operating at room temperatures with device characteristics comparable to silicon based FETs [16,17]. $MoS_2$ belongs to the class of transition metal dichalcogenides (TMDs) consisting of a triple of a transition metal M surrounded by two chalcogenide atoms X with the formula $MX_2$. Particularly, the four stable semiconducting TMDs (SC-TMDs) $MoS_2$, $MoSe_2$, $WS_2$ and $WSe_2$ excel due to their inertness and high stability in ambient conditions similar to graphene or hexagonal boron nitride (hBN), an insulating material with a large band-gap of about ~6 eV [18]. SC-TMDs possess outstanding electronic [6,7,16,19,20], excitonic [21–27], mechanical [28] as well as fascinating spin- and valley [29–32] properties. SC-TMDs feature also various exotic properties including as single photon emission in $WSe_2$ at low temperatures [33–37], photoluminescence efficiency with near unity quantum yield achieved by chemical treatment of sulfur-based SC-TMDs [38,39], layer-dependent superconductivity in highly doped $MoS_2$ [40–44] to name just a view. In addition, more technology relevant characteristics are outstanding FET performance [16], high sunlight absorbance of up to 15% in the visible range [45–47], ultrahigh photo-sensitivity [48–50], catalytic activity [51,52] and photocatalytic stability in aqueous electrolytes [53] making the material not only interesting to replace silicon in electronics but also as photo-catalyst e.g. for solar hydrogen evolution [54] or carbon dioxide reduction [55], but also for cheap optical water disinfection [56].

In this review, we focus on optical properties of the stable SC-TMDs and their van der Waals heterostructures with special emphasize on how these materials interact with light. We begin with a recap of the crystal structure, the striking changes of the electronic band structure by changing the number of layers and the most prominent fabrication methods. The fundamental light matter interaction is determined by the materials complex dielectric function. We discuss the complex dielectric function – a tensor entity – of the SC-TMDs in dependence of the number of layer and highlight differences between the four most prominent examples. Similar to graphene, Raman spectroscopy turns out to be a highly versatile, non-destructive and fast tool for studying 2D materials. The Raman active phonon modes of the TMDs are introduced and the phonon fingerprint for number of layers, strain, temperature, doping and defects summarized. Furthermore, we demonstrate, on example of $MoS_2$, the sensitivity of Raman experiments to study the impact of the environment on the properties of 2D crystals. The combination of different SC-TMD monolayers in vertically stacked van der Waals bilayer heterostructures with type-II band alignment offers potential for device applications such as light-emitting diodes [57,58] and solar cells [59,60], but also to study correlation phenomena in dense systems of bosonic quasiparticles – interlayer excitons, where the photo-excited electron resides in one layer spatially separated from the photo-excited hole in the other layer. The reduced overlap of the electron and hole wave functions results in an increased life-time of the excitons and hence in a high-density exciton ensembles that is predicted to exhibit macroscopic quantum phenomena such as a superfluid phase at rather high temperatures of several tens of Kelvins [61].

We review recent advances in the direction of van der Waals heterostructures with SC-TMDs and provide typical Raman and photoluminescence measurements exemplarily on a $MoS_2$/$WSe_2$ van-der Waals bilayer. The review will be closed by a short summary and the perspectives given by the fact that the opto-/electronic properties and light-matter interaction in SC-TMDs are highly tunable by interfacial engineering [62–67], chemical treatment [38,39], strain [68–70], defect engineering [71], doping [40–44,72–75], in field effect devices [76] and in combination with suitable heterostructures [11,13]. This directly leads to an outlook on the great potential of SC-TMDs in several areas bridging fundamental science and future devices applications due to the rich, manifold properties of SC-TMDs.

## 2. Crystal structure and electronic bands of semiconducting TMDs

The TMD family forms a vast group of more than 60 different compounds that have been studied in the bulk and multilayer forms for decades. The research is motivated by their variety of fascinating properties including metallic, semiconducting, insulating, superconducting and magnetic behavior as well as their thermal, mechanical, catalytic and optical properties including excitonic screening [77]. The maybe longest known compound is $MoS_2$ a.k.a. molybdenite, similar to graphite a natural occurring mineral that can be mined. Due to its layered structure $MoS_2$ is used as dry lubricant [78], but also of interest for e.g. catalysis [51,79] and photovoltaics [80]. The observation of the exciting change of the physical properties with the number of layer [14,15,34,81] were the advent of a recent, intense, and worldwide research on TMDs.

### Crystal structure

TMDs are layered materials with a triple X-M-X layer as smallest unit, where the transition metal M is surrounded by two chalcogen atoms X with strong covalent in-plane bonds, whereas the individual layers hold together by weak van der Waals interaction. The metal atoms are condensed either in octahedral or trigonal prismatic coordination resulting in semiconducting or metallic behavior [19]. One individual layer can be seen as hexagonally arranged $MX_2$ molecules [82]. Different stacking sequence in multilayer structures result in a variety of polytypes with diverse properties. Here we focus on the semiconducting polytype of $MoS_2$, $WS_2$, $MoS_2$ and $WSe_2$ that condenses in 2H-configuration in ABAB stacking sequence with the second layer superimposed to the first one but rotated by 180° around the c-axis as depicted in Fig. 1 (a) [19]. The unit cell of a single layer consists of one $MX_2$ molecule and hence 3 atoms. The lattice symmetry of a single layer is $D_{3h}$ [83] without an inversion center as visualized in Fig. 1 (a). The unit cell of bulk materials consists of two $MX_2$ molecules and hence 6 atoms [compare Fig. 1 (a)] and exhibits $D_{6h}$ symmetry with inversion center. Generally, crystals with an odd number of layers have $D_{3h}$ symmetry and crystals with an even number of layers have $D_{3d}$ symmetry but are often treated like bulk material that has $D_{6h}$ symmetry [83].

Already in 1923 Dickinson and Pauling studied the crystal structure of $MoS_2$ by X-ray diffraction and determined the lattice constants of bulk material to $a$ = 3.15 Å and $c$ = 12.30 Å [84] in the 2D plane and perpendicular to it, respectively. The distance between Mo and S atoms constitutes $d_{Mo-S}$ = 2.42 Å and between two sulfur atoms $d_{S-S}$ = 3.17 Å [85]. The interlayer distance is given by $d$ = 6.5 Å and it is commonly regarded as the thickness of one single layer [19].

### Electronic structure

The electronic band structures of the four SC-TMDs in the focus of this review exhibit very similar signatures with a remarkable dependence on the number of layers due to quantum confinement [14]. The Bloch states defining the single particle band structure are defined by the electron configuration of the atoms together with the bonding structure. In Fig. 1 (c) the band structure determined from density functional theory calculation (DFT) is plotted for bulk, bilayer and monolayer $MoS_2$ [15]. Bulk and bilayer $MoS_2$ exhibit an indirect band gap close to the $\Gamma$-point that is increased by decreasing the number of layers. Nearby is a direct transition located both at the *K*- and the *K'*-point that remain almost unaffected by changing the number of layers. The transitions at the *K*- and *K'*-points are the lowest energy transitions for a monolayer $MoS_2$, because the valence band maximum at the indirect transition is further lowered and simultaneously the conduction band minimum increased so that the indirect transition holds a larger gap compared to the direct transition in the monolayer limit. The transition from an indirect to a direct semiconductor by reducing the number of layers from a bi-layer to a monolayer manifests in a significant increase in the photoluminescence quantum efficiency by several orders of magnitude for mono-layers compared to bi- and multi-layers [14,15].

The origin for this transition is explained by changes in the hybridization between chalcogen $p_z$-orbitals and transition metal $d$-orbitals together with quantum confinement phenomena. The valence band and conduction band states near the $\Gamma$-point are predominantly formed by a linear combination of transition metal $d$-orbitals and antibonding $p_z$-orbitals from the chalcogen atoms. As a consequence these states are significantly affected by interlayer interaction as well as quantum confinement [86]. In turn, the energy of the conduction band is increased near the $\Gamma$-point, whereas the energy of the valence band is decreased with lowering the number of layers [15]. In contrast, both, conduction and valence band at the $K$- and $K'$-points are dominated by localized transition metal $d$-orbitals that lie in the middle of each X-M-X triple unit. Therefore, the energies of the electronic bands at the $K$- and $K'$-points are almost unaffected by interlayer coupling effects and therefore mostly independent from the number of layers. The excitonic transitions at the $K$- and $K'$- points are much brighter in optical emission experiments compared to the transition at the indirect transitions. This qualitative description holds for all SC-TMDs [19,86]. The bandgaps for monolayers MoS$_2$, WS$_2$, MoSe$_2$, and WSe$_2$ constitute 1.8 – 1.9 eV, 1.8 – 2.1 eV, 1.5 – 1.6 eV, and 1.6 – 1.7 eV, respectively [12]. The single particle electronic bands as well as excitonic effects are highly sensitive to environmental and doping induced screening and renormalization effects [23,41,72,75,87–89] and they can be also altered by strain [69,70,90,91] which enables to engineer the opto-/electronic properties.

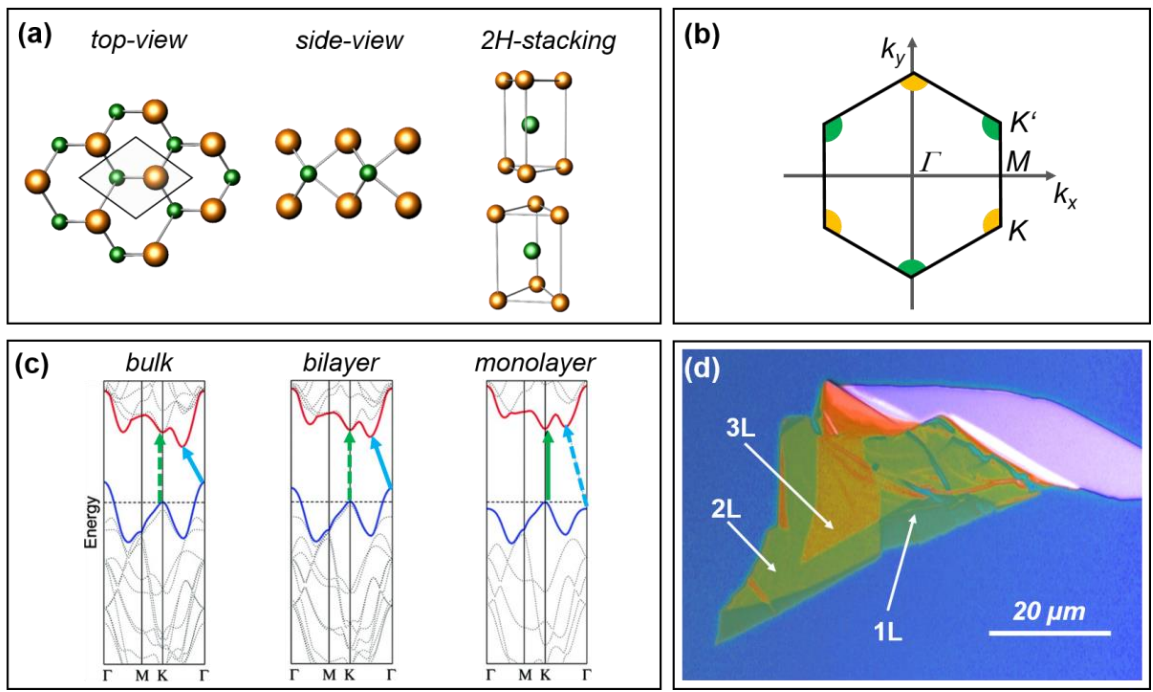

*Figure 1:. (b) Schematic top- and side-view of a monolayer transition metal dichalcogenide (MX$_2$) crystal in 2H phase. The smallest unit is a triple layer, and the atoms are arranged hexagonally in-plane. Stacking order in 2H phase: a monolayer belongs to the $D_{3h}$ point group, a bilayer to the $D_{3d}$ point group and bulk material to the $D^4_{6h}$ point group. (b) Sketch of the hexagonal Brillouin zone of a monolayer MoS$_2$ with the most important symmetry points K, M, $\Gamma$. (c) Calculated band structure of bulk, bi- and mono-layer MoS$_2$ without spin-orbit coupling effects. The lowest energy transitions at the K point (direct) and indirect (close to $\Gamma$-point) are indicated. By lowering the number of layers from bi-to monolayers, the direct transition becomes the lowest energy transition in the band structure causing the transition from an indirect to a direct semiconductor in the monolayer limit. [Adapted with permission from [15]. Copyright 2010, American Chemical Society]. (d) False color representation of an optical micrograph of a MoS$_2$ flake on Si/SiO$_2$ substrate with mono- (1L), bi- (2L), tri-layer (3L) and bulk regions.*

Generally, the *d*-orbitals of the transition metal cause strong spin-orbit coupling (SOC). As a direct consequence, the valence band is spin-spitted due to SOC by ~150 meV for the MoX$_2$ and by ~450meV for the WX$_2$ compounds [12,46,92,93]. The two spin-split optical transitions are called A- and B- excitonic transitions in optical emission and absorption spectra. The band-structure for the monolayer in close vicinity of the *K*- and *K'*- valley can be modelled by an effective 2D massive Dirac Hamiltonian consisting of two additional terms taking into account the finite band gap due to broken symmetry and SOC due to the transition metal *d*-orbitals [93]. The two degenerate high symmetry points in the Brillouin Zone at the *K*- and *K'*- valley and the lack of an inversion center for monolayers locks the spin and valley pseudospin degree of freedom [29,30,32,92,93]. The valley pseudospin can be controlled by excitation with circularly polarized light resulting in interesting valley- and spintronic studies [30,92] such as valley-polarization [29,31] and the observation of the valley Hall effect [32].

Fabrication

For fundamental studies, mono- and few-layer TMD flakes are mostly prepared by micromechanical cleavage from bulk crystals using an adhesive tape [2,3]. The flakes are then transferred to a desired flat substrate material, e.g. Si/SiO$_2$, sapphire or glass. The thin crystals can then be identified by optical microscopy utilizing light interference contrast similar to graphene [94–98]. An example for an optical micrograph of a MoS$_2$ flake deposited on Si/SiO$_2$ substrate with 300 nm SiO$_2$ is shown in Fig 1 (d). The contrast is highlighted by a false color representation demonstrating that mono-, bi-, tri- and few-layer region are clearly distinguishable. This exfoliation method is applicable for almost all 2D materials existing as bulk crystals and results in high-quality and ultra-pure sample perfectly suited for fundamental studies. Exfoliated or grown flakes can be precisely positioned on a substrate using several transfer methods for van der Waals assembly [11,13] e.g. by colamination [18,99,100], PDMS stamping [101] or by a pick and place method [102,103]. The various stacking methods also allow the preparation of vertically stacked van der Waals heterostructures with precise control over the individual layers as well as the rotational alignment [104]. However, the micromechanical cleavage method is not scalable, the yield on large single layer flake is small and lacks control over size and thickness of the flakes.

On the part of scalable fabrication methods, there is great development in both bottom-up and top-down approaches to produce large amounts of high-quality 2D materials for different applications [19,105]. Two typical top-down approaches use bulk materials or powder for liquid exfoliation in solvents [106] or for chemical delamination by chemical intercalation [107] or galvanostatic discharging [108] to produce solutions with a large quantity of TMD nanosheets. Thin films can be prepared out of such solutions e.g. by inkjet printing, spray or dip-coating and filtration [105]. These approaches are scalable and the achieved nano-meshes are suitable for some application. However, these methods do not produce large ultra-pure single crystalline materials.

The synthesis of large area – wafer scale and uniform layers of single layer TMDs is approached by bottom-up methods such as by chemical vapor deposition (CVD) [109–112], metal organic chemical vapor deposition (MOCVD) [113], physical vapor phase growth (PVG) [114,115] or dip-coating of a precursor and thermolysis crystallization by annealing in sulphur gas [116]. There is great progress and it is possible to grow single layer high quality flakes with up to 300 µm in size covering a cm-sized substrate surface [117]. It is still challenging to avoid large gradients across the substrate and to grow homogenously connected single layer films with high crystal quality [19]. Nevertheless, lateral and vertical van der Waals heterostructures with SC-TMDs that are optically and electrically active have been realized by means of CVD [118–121] proving the realistic potential to overcome the scalability issue by the fabrication of single layer SC-TMDs and their heterostructures for future applications.

## 3. *Complex dielectric function and absorbance of TMDs*

The optical properties including absorption efficiency, optical transitions as well as excitonic phenomena are of particular interest for fundamental studies and key to optoelectronic applications. SC-TMDs exhibit strong light matter interaction particularly in the visible range [26]. The light-matter interaction is basically described by the complex dielectric function of a material that helps to link experimental observations and theory to interband excitations. The dielectric function is a tensor entity and therefore the light matter interaction for light field parallel and perpendicular to a 2D crystal is highly anisotropic [47]. The optical properties of SC-TMDs strongly depend on the number of layers [14,15]. For this reason, detailed knowledge of the complex dielectric functions in dependence of the number of layers is highly desirable. Typical methods to extract the dielectric function are ellipsometry measurements under a certain angle of incidence to probe in-plane and out-of-plane components of the light matter interaction done with macroscopically large light spots [122–124]. The fabrication of flakes with a precise control over the number of layers is thus far only achievable by micron sized mechanically exfoliated flakes. Imaging ellipsometry (IE) offers a high lateral resolution while maintaining the precise control over the angel of incidence. IE has been proven to allow access to the dielectric function of 2D materials such as graphene [125], graphene oxide [126] and mono- and multilayer $MoS_2$ [47]. Particularly for monolayer flakes with a thickness of less than 1 nm it can be assumed that interaction with a light-field perpendicular to the 2D sheet can be neglected. Focused reflectivity measurements with the light-field parallel to the 2D plane are suitable to extract the light-matter interaction of SC-TMDs utilizing a Kramers-Kroning constrained variational analysis [46].

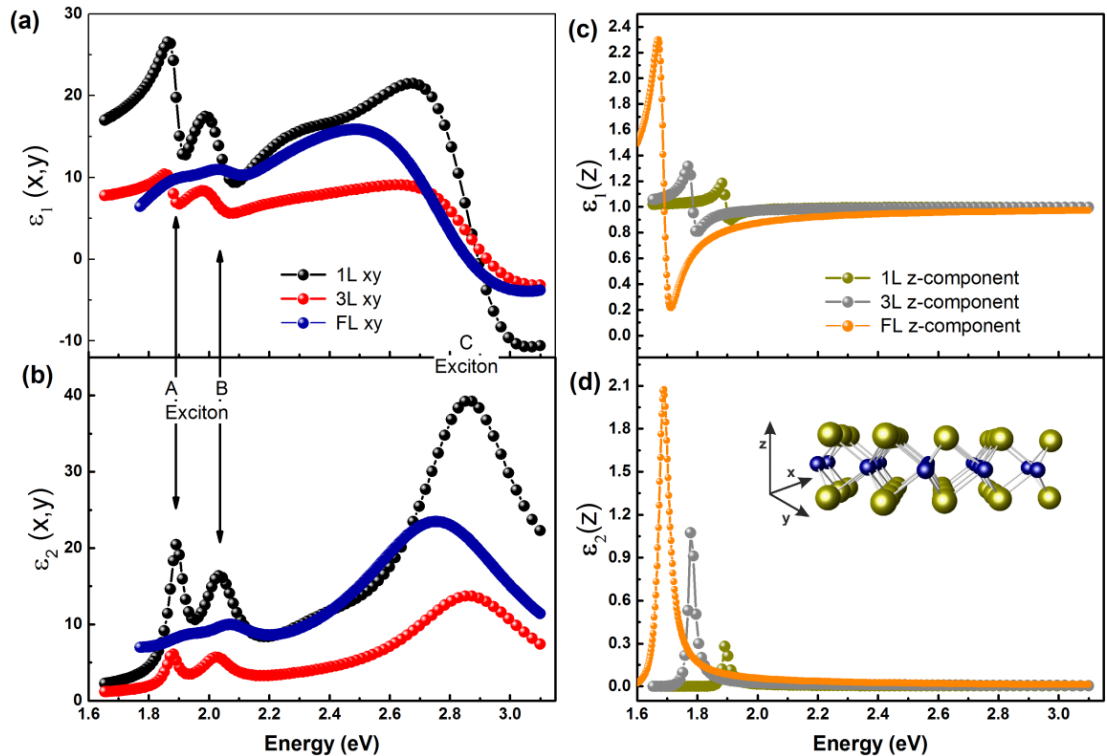

*Figure 2: Complex dielectric function and absorbance of MoS$_2$ from spectroscopic ellipsometry measurements. (a) Real part of the in-plane component of the dielectric tensor $\varepsilon_1(x,y)$ from an anisotropic model for mono-, tri-, four- and few-layer MoS$_2$. (b) Imaginary part of the in-plane component of the dielectric tensor $\varepsilon_2(x,y)$ as in (a) and the data from an isotropic model for comparison. (c) Real part $\varepsilon_1(z)$ and (d) imaginary part $\varepsilon_2(z)$ of the dielectric tensor in out-of plane direction from the anisotropic model to the ellipsometry measurements. Adapted with permission from [48]. Copyright 2016, IOP publishing.*

Layer dependent anisotropic dielectric function and absorbance of $MoS_2$

In a recent study, we investigated the complex dielectric function of mechanically exfoliated mono- tri- and few-layer $MoS_2$ on a transparent sapphire substrate by imaging ellipsometry in a spectral range from 1.7 eV – 3.1 eV. The used angle of incidence of 50° allows to access the optical response of $MoS_2$ parallel (*x,y*) and perpendicular (*z*) to the plane of the flake. A lateral resolution of better than 2 μm enables to probe ultra-high quality crystalline regions. The complex dielectric function for different number of layers have been extracted from the measured ellipsometric angles $\Delta$ and $\psi$ as input to a multilayer model using Lorentz as well as Tauc-Lorentz fit approaches[47]. In the model, the thickness of the $MoS_2$ monolayer was kept at the theoretical value of *d* = 6.15 Å [77,84]. The resulting anisotropic complex dielectric functions $\varepsilon_x=\varepsilon_y\neq\varepsilon_z$ with real part $\varepsilon_1(E)$ and imaginary part $\varepsilon_2(E)$ are plotted for mono-, tri- and few-layer $MoS_2$ in Fig.

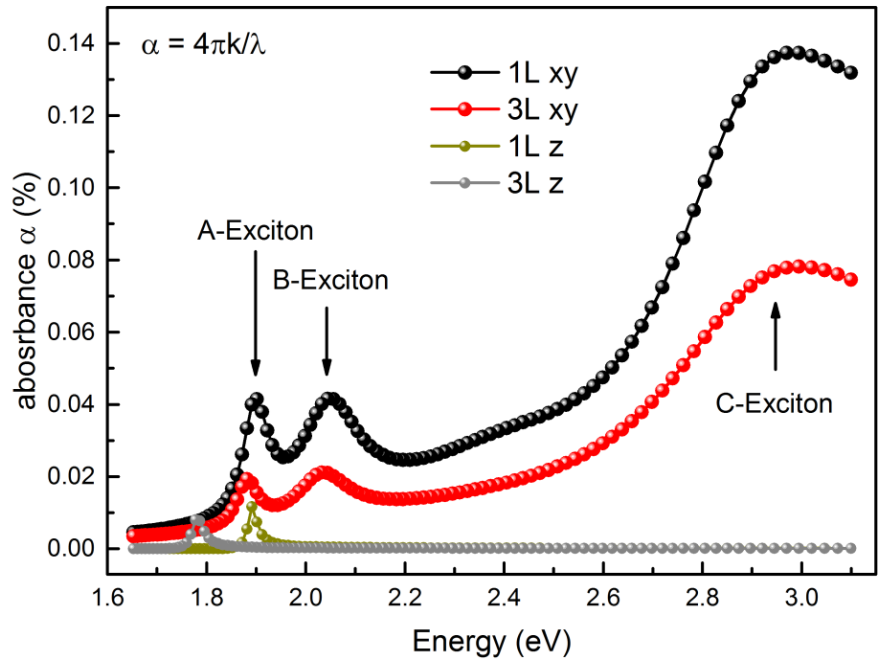

*Figure 3: (e) Absorbance $\alpha$ for mono- and trilayer $MoS_2$ extracted from the optical constants, i.e. from the extinction coefficient in the visible range displaying high absorption efficiencies at A, B and C exciton transitions. The solid spheres are the results deduced from the in-plane component (xy-plane) from the anisotropic model (black mono-layer and red trilayer). The gray and green spheres represent the absorbance perpendicular (z-direction) to the 2D crystal determined from the anisotropic model (mono-layer green, tri-layer gray). Adapted with permission from [48]. Copyright 2016, IOP publishing.*

2 [47]. The in-plane component of the dielectric function $\varepsilon_x(E)=\varepsilon_y(E)$ is best described by 5 Lorentz profiles and the out-of-plane component $\varepsilon_z(E)$ by one Tauc-Lorentz profile, respectively. As visualized in Fig 2 (a, b) the in-plane component exhibits three well pronounced critical points and two rather sharp peaks centered at 1.9 eV and 2.05 eV assigned to the A and B spin-split excitonic transition at the *K-* and *K'*-points in the Brillouin zone. The prominent but rather broad resonance at around 2.9 eV is called C excitonic transition and indicates a high joint density of states [22,77]. This energy range comprises a combination of several individual bright and dark excitonic transitions between 2.2 eV and 3 eV that are located nearby the *Γ-* point [22] in addition to a band nesting region - a region with nearly parallel conduction and valence bands between *M-* and *Γ-* points with a high joint density of states. The individual transitions are significantly broadened by interaction of electrons with optical phonons [22].

The out-of-plane component plotted in Fig. 2(c,d) exhibits one critical point that is located for the monolayer $MoS_2$ at 1.9 eV and hence slightly above the A excitonic transition at the direct band gap. The energy of the critical point is reduced in energy for increasing number of layers to 1.7 eV for a tri-layer and to 1.67 eV for few-layer flake. Consequently, this optical transition is below the direct optical transition at the *K-* and *K'-* points. Therefore, it is most likely related to the states in the Brillouin zone that form the indirect band gap in agreement with the trend that the band gap is reduced for increasing number of layers. The magnitude of the out-of-plane contribution of the dielectric functions is significantly reduced in amplitude compared to the in-plane component and seems to be inversely proportional to the number of layers. This behavior is intuitively clear since the interaction of a light field perpendicular to the 2D film is much stronger for a thicker film compared to an atomically thin mono-layer. Independent from the number of layers, the dielectric functions of the out-of-plane component approach constant values of $\varepsilon_1(z) \approx 1$ and $\varepsilon_2(z) \approx 0$ above approx. 2 eV [47].

The absorbance $\alpha$ of the isolated $MoS_2$ mono-, tri-, and few-layer films can be directly calculated from the dielectric function, taking the relation between the extinction coefficient $\kappa = 2\varepsilon_1\varepsilon_2$ and the absorbance $\alpha = 4\pi\kappa/\lambda$ at a light wavelength $\lambda$. The in-plane and out-of-plane absorbance for mono- and tri -layer $MoS_2$ is plotted in Fig. 3. A monolayer $MoS_2$ absorbs about 4% of the parallel incoming light field at the energy of the A- and B-exciton and amsot 15 % at the C-exciton with a thickness of less than 1 nm [14,45–47]. In the whole visible range, the absorbance is reduced by increasing the number of layers. Particularly, the excitonic resonance at the A- and B-transitions are broadened and reduced in intensity indicating an increased dielectric screening [127]. While the energy positions of the A- and B-excitons are less affected by the number of layers, the energy position of the C-resonance shifts to lower energies by increasing the number of layers, demonstrating that the electronic band structure as well as many body interaction such as the Coulomb-interaction responsible for the exciton formation are located at different *k*-values in the Brillouin zone. These effects are unequally affected by changing the number of layers depending on the atomic orbitals dominating the Bloch states.

The absorbance of a light field perpendicular to the 2D film is 0.2% at the excitonic resonance at 1.9 eV and zero elsewhere. Only 1% of an out of plane light field is absorbed by the excitonic peak for a tri-layer and about 4% for a few-layer $MoS_2$.

The described characteristics of the in-plane dielectric functions $\varepsilon_x(E)=\varepsilon_y(E)$ are in good agreement with the absorbance spectra for exfoliated and CVD grown ultrathin $MoS_2$ layers determined either by spectroscopic ellipsometry with isotropic or anisotropic fit approaches or extracted from reflectance measurements [46,47,122,123,128,129]. The universality of the main characteristics of the dielectric functions underlines the robustness of the strong light-matter interaction in atomically thin MoS$_2$ layers. However, we would like to emphasize that the strength as well as the specific spectral position of the A-, B- and C-excitonic transitions are unique for each sample and highly affected by the individual situation such as environment, substrate, doping or strain just to name some major sources modifying the light-matter interaction. As described above, the different critical points in the electronic band structure with a high joint density of states are formed by different atomic orbitals of the transition metal and sulfur atoms, respectively. The A- and B-excitons can consequently be affected differently by perturbations and changes compared to the higher energy transitions at the C-excitonic resonance or to the spectral region between B- and C-exciton transitions.

Comparison of in-plane dielectric function and absorbance for MS$_2$ and XSe$_2$ (M = Mo, W)

All four SC-TMDs are characterized by strong light-matter interaction resulting in similar values and overall characteristics in absorbance spectra as well as dielectric functions in the visible range as demonstrated in Fig. 4 for MoSe$_2$, WSe$_2$, MoS$_2$, and WS$_2$ monolayers [46]. The 2D crystals have been prepared from bulk crystals by micromechanical exfoliation and placed on fused silica substrates. The optical parameters have been extracted from reflection measurements with the incident focused light beam perpendicular to the 2D plane. The light-field in such a configuration is parallel to the sample plane and therefore only the in-plane light matter interaction can be probed. The in-plane components of the dielectric functions $\varepsilon_x(E) = \varepsilon_y(E)$ are derived from the reflection spectra by a Kramers-Kroning constrained analysis [46].

As already introduced above, the absorbance can be directly extracted from the dielectric functions. The absorbance spectra for quasi free-standing and supported monolayers are displayed in Fig. 4 (a, b), respectively. The authors of Ref. [46] have calculated the absorbance spectra for the supported monolayers starting from the identical dielectric function [shown in Fig. 4 (c, d)], but taking into account a local field correction factor for the light intensity above the substrate that depends on the refractive index of the substrate material [46]. The absorbance of the free standing material constitutes in the visible range up to 15% and is reduced by about 1/3 for the monolayers on fused silica [46]. The high absorbance values underline the strong light matter interaction of SC-TMDs even in the monolayer limit with a thickness of less than 1 nm as it has been reported by several theoretical and experimental studies, too [14,22,26,27,45–48,122,124,128,130]. The sizeable reduction of the absorbance for monolayers on a substrate points to the fact that the optical response can be significantly altered just by modification of substrate or environment by dielectric engineering and screening effects [38,62,63,65,74,89,127,131–133]. The strong influence of the environment on the light-matter interaction of atomically thin semiconducting membranes motivates a great potential not only for sensing applications, but also for novel device architectures with precisely tailorable optical properties.

The critical points in the optical response found in the complex dielectric functions and absorbance spectra are similar for MoSe$_2$, WSe$_2$, MoS$_2$, and WS$_2$ monolayers. The lowest energy transitions labeled as A- and

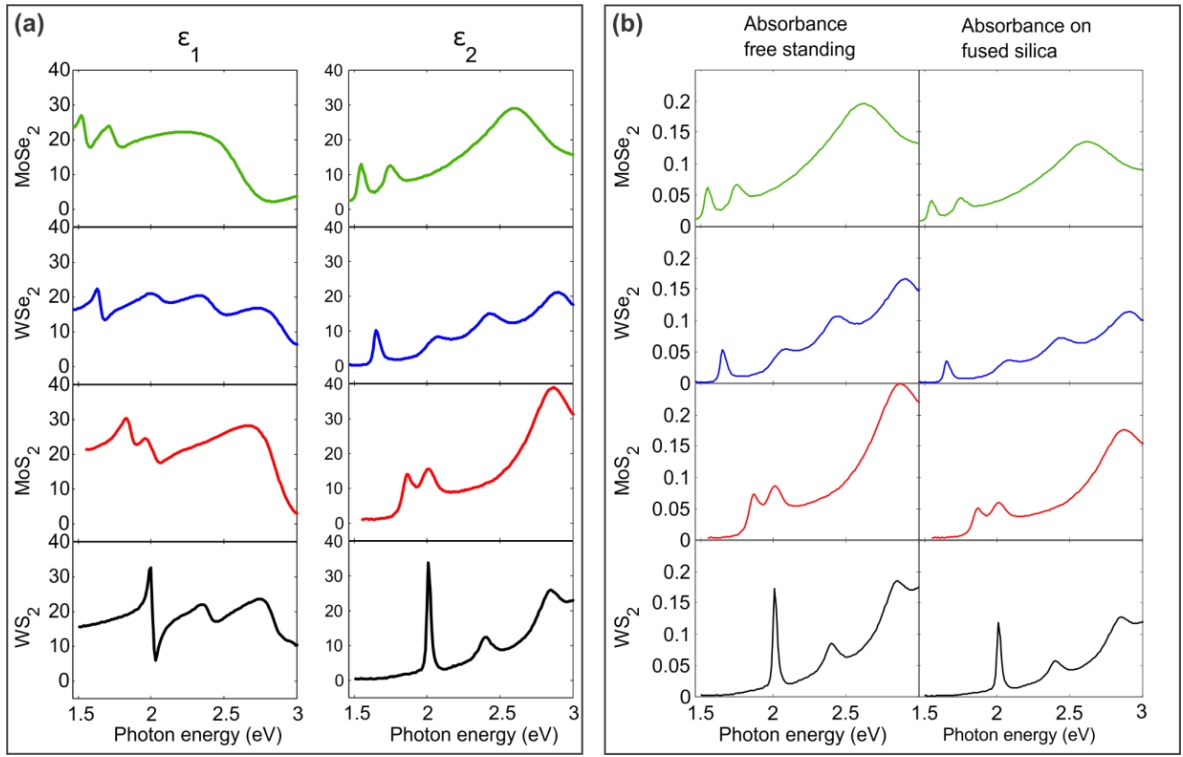

*Figure 4: Absorbance for free standing (a) and supported on fused silica (b) micro-mechanically exfoliated MoSe$_2$, WSe$_2$, MoS$_2$, and WS$_2$ monolayers [47]. In plane components of the real part $\varepsilon_1$ (c) and imaginary part $\varepsilon_2$ (d) of the dielectric functions determined from reflection measurements for ) micro-mechanically exfoliated MoSe$_2$, WSe$_2$, MoS$_2$, and WS$_2$ monolayers on fused silica [47]. All panels Reprinted figures with permission from [47]. Copyright 2014 by the American Physical Society.*

B-exciton belong to the direct optical transitions at the *K-* and *K'-* points of the band structure. The energy of the A-exciton is the single particle band gap minus the exciton binding energy which is in the order of several hundreds of meV [22,25]. Since both, the single particle band structure and the Coulomb interaction determining the exciton binding energy are reduced by screening effects such as e.g. doping [23,72,73,75], the A-excitonic transition energy is only minor affected by screening. However, the oscillator strength in both absorption and photoluminescence is significantly altered by screening and doping [63,72,87,89,131]. The B-exciton belongs to the *K-, K'-* points in momentum space and is the transition from the lower energy spin split valence band state to the conduction band [92]. Since the Coulomb interaction is comparable for A- and B-exciton, the energy splitting between A- and B-exciton in the absorbance is a direct measure for the spin-splitting in the valence band due to spin-orbit coupling. The spin-splitting of the valence bands at the *K-* and *K'-*points constitutes between 150 meV and 200 meV and between 400 and 450 meV for the MoX$_2$ and WX$_2$ (X = S, Se) monolayers, respectively. The increase in the spin splitting of the tungsten based compounds is due to the larger mass of tungsten and hence, an increase in the spin-orbit coupling strength of the tungsten *d*-orbitals. Similar to the discussion of the broad C-resonances for MoS$_2$ monolayers above, also the broad C-excitonic signatures for the other compounds belong to higher lying interband transitions with a high joint density of states at different areas in momentum space. The same holds for the so-called D-resonance observable for WSe$_2$ in the plotted spectral range [46].

MoSe$_2$, WSe$_2$, MoS$_2$, and WS$_2$ monolayers exhibit superior light-matter interaction properties and exciton dominated optical response in the visible range with maxima in the peak absorbance for freestanding materials in the order of 15% for less than 1 nm thick crystals. As expected for 2D materials, the optical responses described by the complex dielectric functions are highly anisotropic for the interaction of light field parallel or orthogonal to the plane of the 2D materials. The light-matter interaction is reduced for increasing number of layers and the excitonic signatures are broadened presumably due to a decreased Coulomb interaction and hence decreased excitonic phenomena. However, also further decay and relaxation channels of the carriers by interlayer scattering exist. Particularly, the C-excitonic transition is redshifted with an increasing number of layers. The same observation holds for the only critical point in the out-of-plane component of the dielectric function $\varepsilon_z(E)$. It is found that this resonance is redshifted with increasing number of layers. The amplitude however is enlarged with the number of layers as intuitively expected since the thickness in *z* direction increases.

## 4. *First order phonon modes as unique fingerprint for material properties*

Raman spectroscopy is a very powerful tool to study many structural but also electronic properties of 2D materials due to the sensitivity of the fundamental Raman active lattice vibrations to perturbations of the lattice but also to changes in the electronic properties caused by electron-phonon interaction and phonon renormalization effects as well as more indirectly, by resonance effects if the incoming or scattered light meets a fundamental optical interband transition of the electronic band structure. Raman scattering is a versatile, non-destructive fast characterization tool without the need for extensive device fabrication. Its powerfulness has significantly supported the success of graphene research [134,135]. The strength of Raman investigations continues for research on MoS$_2$ and other TMDs. However, there is one fundamental difference between graphene and SC-TMDs: due to the gapless linear band structure of graphene Raman spectroscopy is resonant for all excitation energies enabling e.g. the activation of the double resonant 2D mode in graphene [134,135]. SC-TMDs hold a finite band-gap and therefore resonant Raman spectroscopy is only possible if the incoming or scattered light meets a fundamental optically active interband transition [134,136]. In this respect, resonant Raman spectroscopy with varying photon energies provides a complementary tool to emission and absorption measurements to access the electronic structure in SC-TMDs as well as to study fundamental aspects of the electron-phonon interaction [134,136,137]. Already non-resonant Raman spectroscopy, meaning that neither the exciting (incoming resonance) nor the scattered photon energy (outgoing resonance) meet an optically active interband transition of the electronic bands, provides access to the number of layers [81,138–142], phase transition 2H, 3R, 1T, 1T' MX$_2$ crystals [137,143,144], strain [69,145,146], disorder and defects [147], doping [74,148] and temperature as well as thermal conductivity [97,149–152]. Amazingly, this information can already be extracted by investigating only two Raman active zone center phonon modes in the higher frequency range ($\omega_{ph}$ > 200 cm$^{-1}$). These are an in-plane mode with $E^1_{2g}$ *(E')* symmetry and a homopolar out-of-plane mode with $A_{1g}$ *(A'$_1$)* symmetry for bulk (mono-layer) crystals. The in-plane mode is the degenerate LO/TO phonon of the material. Overall, 2H-TMDs possess four Raman active phonon modes with the displacement of the atoms sketched in Fig. 5 (a). Besides the two high frequency modes, there are two lower frequency modes ($\omega_{ph}$ > 45 cm$^{-1}$), the layer breathing mode and the interlayer shear mode with $E_{1g}$ *(E'')* and $E^2_{2g}$ symmetry, respectively. Obviously, for monolayers only the breathing mode with *E''* symmetry exist.

Particularly, the interlayer shear mode has proven as a very sensitive measure for the interlayer coupling in artificially stacked bi- and multilayers [153] as well as for the dependence of the interlayer coupling strength from the twist angel between two monolayers [154]. Not only for monolithic devices but also for van der Waals hetero bi- and multilayers, the interlayer coupling strength can be evaluated by the energy of the interlayer shear and breathing modes [155]. Optical signatures for van der Waals heterostructures and the interlayer coupling strength will be discussed in details in a separate section below.

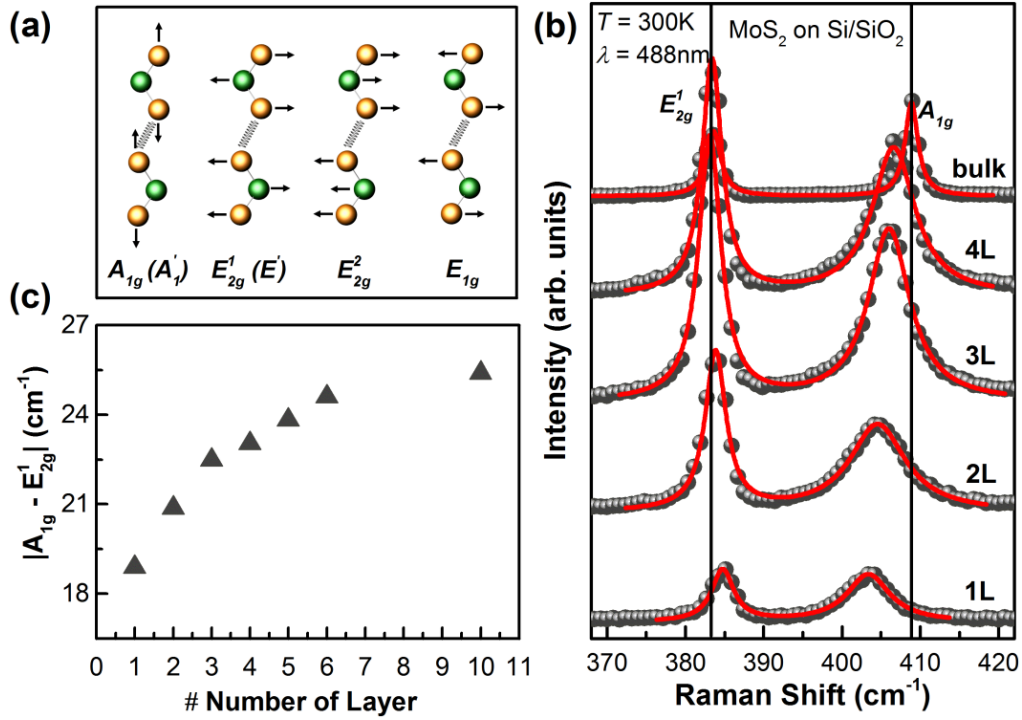

*Figure 5: (a) Visualization of the displacement of the atoms for the Raman active phonon modes in TMDs. (b) Raman spectra for micromechanically exfoliated $MoS_2$ mono-, bi-, tri-, four-, layer and bulk flakes. The solid lines are fits to the spectra with 2 Lorentzian ($MoS_2$ on $Si/SiO_2$, excited with $\lambda$ = 488 nm). (c) Energy difference between the two Raman active phonon modes $A_{1g}$ and $E^1_{2g}$ from measured for $MoS_2$ from (a) that is an explicit measure to determine the number of layers.*

In the course of this section, we focus on aspects of non-resonant Raman investigations of the higher frequency intra-layer Raman active modes of SC-TMDs with special emphasize on the determination of the number of layers, on doping induced phonon-renormalization and the resulting options to study the change in doping density without the needs of contacts and leads. Furthermore, we cover the impact of strain, defects and temperature on the phonon modes. The Raman tensor of the two modes of interest are for the $A_{1g}$ ($A'_{1g}$) symmetries $\begin{pmatrix} a & 0 & 0 \\ 0 & a & 0 \\ 0 & 0 & b \end{pmatrix}$ and for the $E^1_{2g}$ ($E'$) symmetries $\begin{pmatrix} 0 & d & 0 \\ d & 0 & 0 \\ 0 & 0 & 0 \end{pmatrix}, \begin{pmatrix} d & 0 & 0 \\ 0 & -d & 0 \\ 0 & 0 & 0 \end{pmatrix}$.

As a result, the out-of-plane mode is for linearly and circularly polarized light observable in co-polarization geometry and the in-plane mode is un-polarized for linearly polarized light and co-polarized for circularly polarized light [156]. For a more detailed description of the classical as well as quantum description of the Raman tensor of TMDs, the inelastic light scattering itself and consequence for the polarization dependence we would like to refer to a recent comprehensive review by Saito *et al.* [137].

Counting the number of layers from Raman measurements

As introduced already in 2010 by Lee et al. [81] for MoS$_2$, the frequencies of the first order phonon modes with $E^1_{2g}$ and $A_{1g}$ symmetry change monotonously with the number of layers with opposite trends. Raman spectra taken on exfoliated MoS$_2$ on Si/SiO$_2$ substrate are plotted in Fig. 5(b). The in-plane mode $E^1_{2g}$ softens and the Raman signal is therefore red-shifted with increasing number of layers, whereas the out-of-plane mode $A_{1g}$ is significantly stiffened and the Raman frequency is blue-shifted. Therefore, the energy difference between the two modes $\Delta E = |\omega(A_{1g})-\omega(E^1_{2g})|$ allows to determine the number of layers with a very high precision as demonstrated in Fig. 5 (c). Particularly mono-, bi and tri-layer can be unambiguously determined.

The stiffening of the out-of-plane $A_{1g}$ mode with increasing number of layers is well understood considering the classical model for coupled harmonic oscillators [81]. The unusual behavior of the in-plane $E^1_{2g}$ mode to soften with increasing number of layers is most likely due to modification of long range Coulomb interlayer interactions induced by the additional layer causing an enhanced dielectric screening between the effective charges in MoS$_2$ [81,142].

|  | MoS$_2$ [81] | WS$_2$ [139] | MoSe$_2$ [141] | WSe$_2$ [141] |
|---|---|---|---|---|
| **E' (1L)** | 384.2 cm$^{-1}$ | 355.2 cm$^{-1}$ | 287.2 cm$^{-1}$ | ≈ 250 cm$^{-1}$ |
| $E^1_{2g}$ *(bulk)* | *382.0 cm$^{-1}$* | *355.3 cm$^{-1}$* | *285.9 cm$^{-1}$* | *248.0 cm$^{-1}$* |
| **A'$_1$ (1L)** | 403.0 cm$^{-1}$ | 417.2 cm$^{-1}$ | 240.5 cm$^{-1}$ | ≈ 250 cm$^{-1}$ |
| **A$_{1g}$ (bulk)** | *407.1 cm$^{-1}$* | *420.1 cm$^{-1}$* | *242.5 cm$^{-1}$* | *250.8 cm$^{-1}$* |

Table 1: Energies for the most prominent first order Raman modes E'/$E^1_{2g}$ and A'$_1$/$A_{1g}$ belonging to the Raman active optical phonon branches of the phonon dispersion at the $\Gamma$-point for MoS$_2$, WS$_2$, MoSe$_2$, and WSe$_2$ monolayers and bulk materials, respectively. Notations are given with respect to the $D_{3h}$/$D_{6h}$ point groups for monolayer/bulk materials condensed in 2H symmetry. All values are extracted from measurements on mechanically exfoliated flakes excited at room temperature with a laser wavelength of $\lambda$ = 514.5 nm.

These two interlayer phonon modes are useful to determine the number of layers for other SC-TMDs, too. In Table 1, the relevant mode frequencies are summarized for monolayers and bulk materials of MoS$_2$, WS$_2$, MoSe$_2$, and WSe$_2$. The same trend as for MoS$_2$ holds for the two interlayer phonon modes of WSe$_2$ and MoSe$_2$. Only for mono-layers of WSe$_2$, the E' and A'$_1$ modes are degenerate with an frequency of $\omega \approx$ 250 cm$^{-1}$ [141] and cannot be separated in experiments. However, they are split for WSe$_2$ bulk and the (degenerate) mode frequencies change with the number of layers so that an estimation of the number of layers is also possible for WSe$_2$ [141].

Phonon renormalization due to doping

Fig 6. (a) and (b) show the E' ($E^1_{2g}$) and A'$_1$($A_{1g}$) phonon mode frequencies for monolayer and bilayer MoS$_2$ as a function of the top gate voltage using a solid electrolyte gate electrode [74]. The flakes have been placed on a highly p-doped Si back electrode covered with 300 nm SiO$_2$ and contacted by 2 Ti/Au contacts. A solid electrolyte gate using polyethylene oxide (PEO) and cesium-perchlorate is utilized because its high capacity allows to tune the charge carrier density over a wide range by applying a rather small voltage [148]. With the top gate, the 2D electron density in the MoS$_2$ flakes can be tuned by two orders of magnitude from approx. 10$^{11}$ to 10$^{13}$ cm$^{-2}$ by applying a top gate voltage of $V_{TG}$ = ± 1V [74]. Raman measurements are

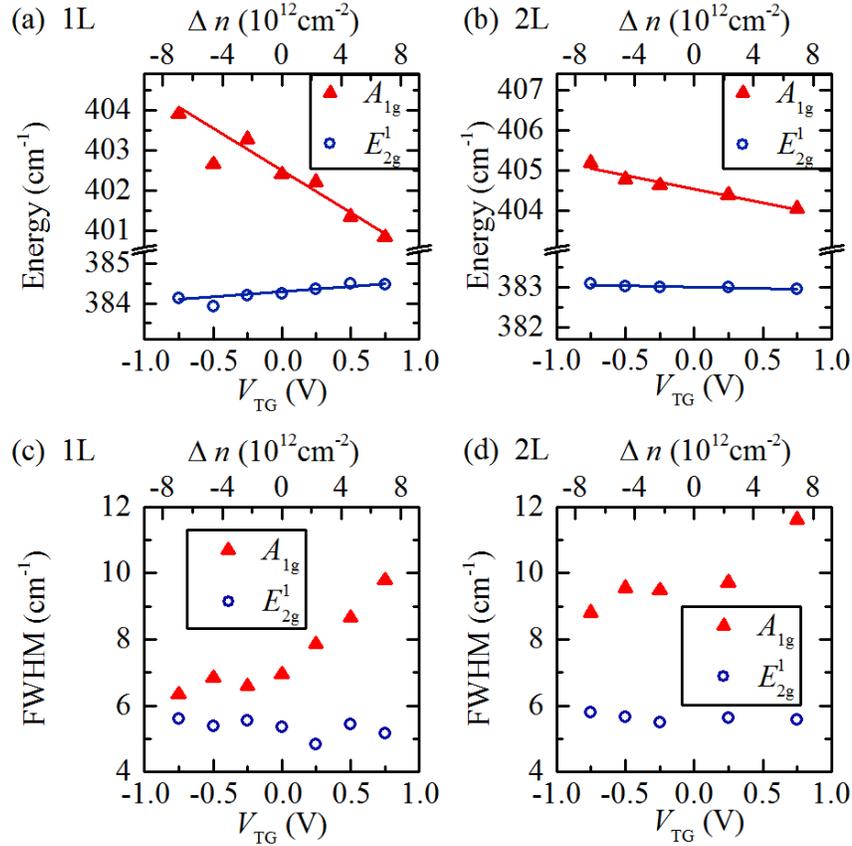

*Figure 6: Energy and full width at half maximum (FWHM) of the $A_{1g}$ and $E^1_{2g}$ modes in dependence of the applied top gate voltage. (a) Phonon mode energies for monolayer MoS$_2$ and (b) for bilayer MoS$_2$. Solid lines are linear fits to the data points. Panel (a, b) Reprinted from [75], with the permission of AIP Publishing 2015. (c) FWHM of the phonon modes extracted from Lorentzian fits to the Raman spectra for mono-layer and (d) for bi-layer MoS2. The Raman measurements are carried out at room temperature with excitation wavelength of $\lambda$ = 488 nm.*

carried out at room temperature with an excitation wavelength of $\lambda$ = 488 nm. For both, mono- and bilayer MoS$_2$ the phonon frequencies of the in-plane modes are almost unaffected by changing the gate voltage and hence the carrier concentration, whereas the out-of-plane modes are redshifted by $\Delta\omega(A') \approx 3$ cm$^{-1}$ for the monolayer and $\Delta\omega(A') \approx 1.5$ cm$^{-1}$ for the bilayer by changing the charge carrier density by approx. two orders of magnitude. In both cases, the change in the mode frequencies seems to be directly proportional to the applied gate voltage and therefore to the 2D charge carrier density. A line-shape analysis using two Lorentz profiles to describe one Raman spectra reveals, that simultaneously the FWHM of the out-of-plane modes increase with increasing charge carrier densities. By increasing the charge carrier density from ~10$^{11}$ to ~10$^{13}$ cm$^{-2}$, the FWHM increases from approx. 6.5 cm$^{-1}$ to 10 cm$^{-1}$ for the monolayer and from 8.8 cm$^{-1}$ to 11.6 cm$^{-1}$ for the bilayers [Figs. 6 (c, d)]. The in-plane modes are again unaffected by changing the charge carrier density and the FWHM constitutes for mono- and bilayers to be approx. 5.5 cm$^{-1}$.

Very similar qualitative and quantitative findings in Raman investigations on gated MoS$_2$ monolayers have been reported in a comparable range of charge carrier densities by Chakraborty *et al.* [148]. The authors calculated the charge carrier dependence of the electron phonon coupling from first-principle functional density theory. They found an excellent agreement between theory and experiments for doped MoS$_2$ monolayers: theoretically the strengthening electron phonon-coupling with increasing charge carrier density results in a significant softening and broadening of the out-of-plane phonon. The in-plane mode is hardly affected by doping [148]. The different reaction of the two optical phonon modes on doping results from their different symmetry. The *A'* (*A$_{1g}$*) mode has shares the symmetry as the lattice and the electronic bands at the *K*- and *K'*- points in momentum space. By doping with electrons, the antibonding states with Mo $d_{z2}$ orbital character are filled which reduces the bond strength. As a result the out-of-plane phonon mode is softened. The occupation of the conduction band minimum states at the *K*- and *K'*- points induces an increase in the electron-phonon coupling [148]. Such a doping induced phonon renormalization is absent for the in-plane mode with *E'* (*E$^1_{2g}$*) symmetry due to its orthogonality to the *A'* (*A$_{1g}$*) symmetry. Therefore, the matrix element that couples electrons and phonons vanishes for the in-plane mode because of this orthogonality [148].

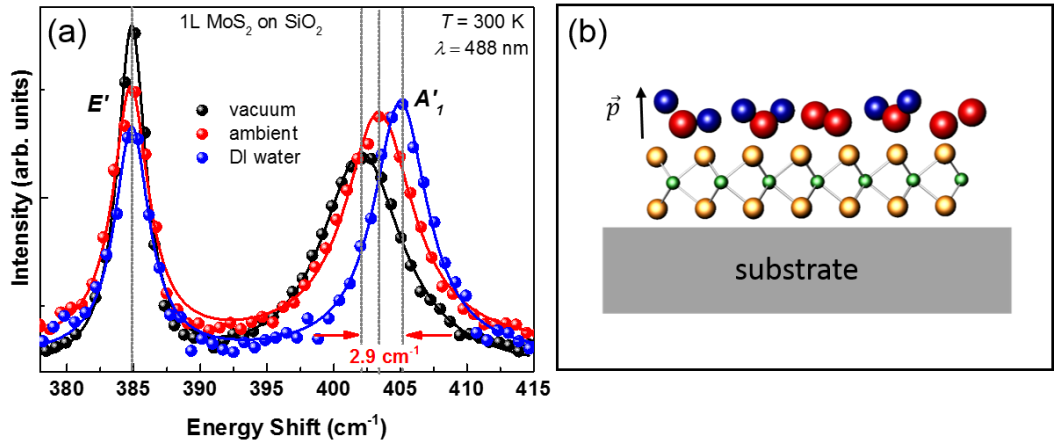

*Figure 7: (a) Environmental dependence on the Raman spectra for micromechanically exfoliated mono-layer of MoS$_2$ on Si/SiO$_2$ substrate measured at room-temperature under non-resonant excitation (λ = 488 nm, P = 100 µW, spot diameter < 1 µm). The measurements are performed in DI-water, ambient and vacuum (p ≤ 5 x 10$^{-6}$ mbar). Filled spheres are the experimental data and solid lines are fits to the data using two Lorentz profiles. (b) Sketch of the MoS$_2$ flake on substrate covered with water molecule causing a finite dipolar moment and hence acting as molecular gates.*

It is interesting to see that the general behavior of the two optical phonon modes is also valid for bilayers [compare Fig. 6 (b,d)] and seemingly also for tri- and four-layer crystals [74] indicating that the conduction band minimum at the *K*- and *K'*- points matters for this thin 2D crystals with an indirect bandgap. The symmetry argument that the phonon-renormalization is only effective for the *A$_{1g}$* modes is valid also for bi- and tri and four-layer MoS$_2$.

Overall, the sensitivity of the out-of-plane phonon mode with *A$_{1g}$* symmetry empowers Raman experiments to identify the charge carrier density in atomically thin SC-TMDs without the need to fabricate ohmic contacts and to apply magnetic fields to conduct Hall measurements – the standard probe to determine the charge carrier density in doped semiconductors. Since the explanation of the doping induced phonon renormalization holds also for other SC-TMDs with identical lattice symmetry and direct band gaps at the *K*- and *K'*-points, the determination of the charge carrier density from Raman measurements is expected to be also suitable for all four TMDs of in the focus of this review.

An example for this convenient method to determine the change in the charge carrier density of monolayer MoS$_2$ is provided by the results shown in Fig. 7 (a). Non-resonant Raman spectra ($\lambda_{laser}$ = 488 nm at room temperature) taken on a micromechanically exfoliated MoS$_2$ monolayer on a Si/SiO$_2$ substrate are contrasted for measurements in vacuum ($p \leq 5 \times 10^{-5}$ mbar), ambient conditions and with the flake immerse in DI-water (water measurements are performed using a water-dipping objective as described elsewhere [53]). The frequency as well as linewidth of the in-plane $E'$- mode is rather constant for all environments. The frequency of the out-of-plane $A'_1$- mode is highest for measurements in water, but reduced by almost 3 cm$^{-1}$ and broadened for spectra taken in vacuum. As shown by us in an earlier work [74], the $A'_1$- mode frequency and its broadening in ambient conditions strongly depends on the laser power and varies between the two limits for measurements in water and vacuum. The mode frequencies for lowest laser power approaches the values measured in water, whereas the one observed with higher laser power are close to the mode frequencies observed in vacuum. The experimental observations are consistent with a change of the charge carrier density. The free electron density for measurements in vacuum is up to two orders of magnitudes higher compared to the measurements with the MoS$_2$ flakes completely immersed in DI water. We attribute this change to the effect of molecular doping. Due to their dipolar moments, some molecules such as e.g. O$_2$ and H$_2$O can act as molecular gates causing the transfer of a fraction of a charge from the 2D material to the molecules [63,87] effectively depleting the 2D electron system in MoS$_2$. By illumination the flakes in vacuum, the surface of the 2D material is cleaned and physisorbed molecules are removed. In vacuum, we assume that the intrinsic charge carrier density is measured. Intrinsically, MoS$_2$ is unintentionally n-type doped with electron densities exceeding 10$^{13}$ cm$^{-2}$ most likely because of a high density of sulfur vacancies [16,24]. Immersed in water, the surface of a MoS2 flake is completely covered with H$_2$O and O$_2$ molecules as sketched in Fig. 7(b) resulting in the most efficient molecular gating effect resulting in a depletion of the MoS$_2$ flake by up to two orders of magnitude. The laser power dependence is explained by a gradual removal of physisorbed molecules with the steady state between adsorption and removal depending on the laser power. Since this process is highly reversible, the charge carrier density of MoS$_2$ in ambient conditions can be tuned just by the illumination intensity [74].

Impact of strain and defects on the phonon modes

Opposite to the doping that effects only the homopolar out-of-plane phonon with $A'_1$ ($A_{1g}$) symmetry, strain predominantly effects the in-plane phonon with $E'$ ($E^1_{2g}$) symmetry [69,145,146]. In particular, uniaxial strain induces a reduction of the mode frequency per % tensile strain of $\Delta\omega/\varepsilon$ = -2.1 cm$^{-1}$ for the $A'_1$, $A_{1g}$ phonon mode and only $\Delta\omega/\varepsilon$ = -0.4 cm$^{-1}$ for the $E'/E^1_{2g}$ phonon mode [145]. For the application of uniaxial tensile strain for monolayers MoS$_2$, no effect on the $A'_1$ mode is reported, however the E'-mode is substantially softened and a mode splitting appears for tensile strain larger than 0.8% [69]. Bi-layer MoS$_2$ is affected in a similar way [69]. The mode splitting of the doubly degenerate $E'$ mode is assigned to symmetry breaking of the crystal caused by uniaxial strain. We would like to mention that the application of strain not only affects the zone center phonon modes but also the electronic band structure. The bandgap is lowered with increasing tensile strain and the strained material can even undergo the transition from an direct to an indirect semiconducting material [69,146].

The effect of disorder on the Raman spectrum of mono-layer MoS$_2$ is reproduced in Fig. 8 [147], where the defects are induced by a bombardment with Mn$^+$ ions with varying dose enabling a systematic study of the Raman response with respect to the interdefect distance $L_D$. The observed changes in the Raman spectra contrasts the modifications in the spectra induced by strain or doping. With increasing defect density, the *E'*-phonon frequency is lowered, and the frequency of the *A'* mode is simultaneously increased [147]. Both modes are broadened with increasing defect density and the LA(M) mode, a longitudinal acoustic mode at M point at the Brillouin zone boundary is activated. A combination of ion bombardment caused doping together with strain cannot explain the experimental signatures since the A' mode frequency is not only increased with increasing ion does but also significantly broadened excluding a reduction of the electron density to be the main source for the modification of the out-of-plane phonon mode. Mignuzzi *et al.* [147] adopted an already established phonon confinement model [157,158] to explain the defect induced changes of the first order phonon modes. The authors find excellent agreement between experiment and numerical calculations of the phonon lineshape utilizing the phonon confinement model together with the phonon-dispersion from density functional theory calculations [147]. Furthermore, the intensity ratios between the defect activated phonon modes such as the LA(M) and the first order *E'* and *A'$_1$* phonons increases monotonically with decreasing interdefect density *L$_D$* [147] serving as a very suitable estimate for the defect density from Raman measurements in addition to the opposite frequency shifts of the *E'* and *A'$_1$* phonons.

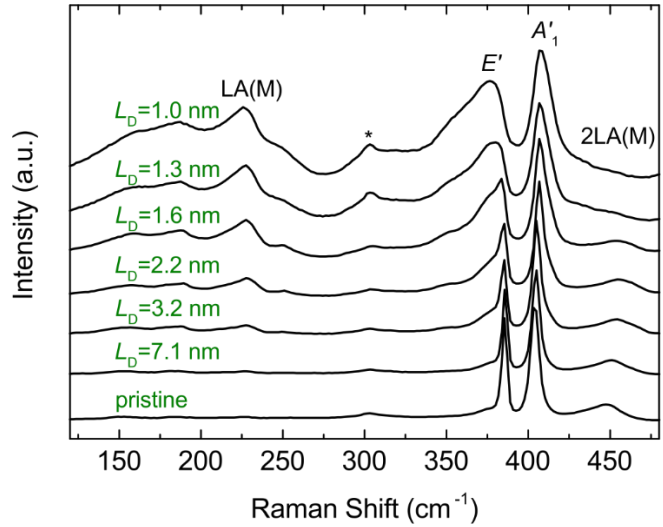

*Figure 8: Raman spectra of a pristine and ion bombarded (Mn+-ions) monolayers of MoS$_2$ on Si/SiO2 with varying interdefect distances L$_D$. The asterisk refers to a Raman peak of the Si substrate. The spectra have been normalized to the intensity of the A'$_1$ mode. The spectra are taken at room temperature in ambient conditions and have been excited using a laser with $\lambda_{laser}$ = 532 nm. Reprinted figure with permission from [148]. Copyright 2015 by the American Physical Society.*

Temperature shift of the phonon modes

Temperature induced changes in the lattice constant of a crystal have immediate consequences on the phonon mode frequencies. Typically, the lattice constant increases with increasing temperature. As a direct consequence the phonon modes are softened. Therefore, Raman measurements provide a fast, non-contact method to monitor local changes in the temperature with the spatial resolution of the laser spot that can be better than 1 µm. A raise in temperature can be directly induced by light irradiation, by electric currents in electronic devices and other heat sources. Raman measurements with a high spatial resolution can not only be utilized to measure the local temperature but also to investigate the thermal conductivity of a material. It has been shown that both optical phonon modes with $E'/E^1_{2g}$ and $A'_1/A_{1g}$ symmetry of SC-TMDs are lowered with increasing temperature [149–152]. The first order temperature coefficients $\chi_T$ for MoS$_2$ are reported to be linear for both phonon modes. The temperature coefficients are reported to be very similar for supported and suspended mono-layers. The temperature coefficients constitute for suspended MoS$_2$ mono-layers $\chi_T$ (E') = -0.011 cm$^{-1}$/K and $\chi_T$ (A'$_1$) = -0.012 cm$^{-1}$/K for the E' and A'$_1$ modes, respectively and for sapphire supported MoS$_2$ monolayers $\chi_T$ (E') = -0.017 cm$^{-1}$/K and $\chi_T$ (A'$_1$) = -0.012 cm$^{-1}$/K [149]. With the help of Raman measurements, the thermal conductivity $\kappa$ of a monolayer MoS$_2$ at room temperature was determined to be $\kappa$ = 34.5 W/mK [149], which is significantly smaller than the thermal conductivity of graphene with $\kappa$ = 2000 W/mK [159].

Overall, Raman investigations provide unique access to both number of layers and doping, as well as strain, defect densities and temperatures. Particularly, for mono-layers and ultrathin SC-TMDs the changes in the high frequency first order Raman modes with $A'_1(A_{1g})$ (out-of-plane) and $E'(E^1_{2g})$ (in-plane) symmetries that are easily experimentally accessible provide unique and unequivocal access to the above mentioned parameters. A temperature increase shifts both modes to lower frequencies, whereas increasing defect densities shifts the in-plane mode to lower frequencies and the out-of-plane mode to higher frequencies and both modes are getting broadened. A decrease in the charge carrier doping also shifts the out-of-plane mode to higher frequencies however the mode is then getting narrower and the in-plane mode is unaffected. These characters of the two higher frequency phonon modes enables to clearly separate between temperature, defect and doping related effects on the Raman response. Moreover, the effect of uni- and biaxial strain on the Raman modes can be clearly distinguished since strain mainly affects the in-plane mode, while the out-of-plane mode is mostly unchanged. With increasing tensile strain, the in-plane mode frequency is reduced, broadened and eventually the degeneracy of the TO and LO phonon branches lifted resulting in a splitting of the in-plane mode.

### 5. *Optical properties of TMDC heterostructures*

Engineering and controlling the optical and optoelectronic properties of semiconducting materials on demand is highly desirable for technological applications, e.g. for solar cells and light emitting but also to design host materials for studying emergent quantum phenomena for instance in dense exciton ensembles. For the realization of optoelectronic devices, bipolar heterostructures are often required. Van der Waals heterostructures can be electrostatically doped by gate structures, because there are intrinsic n-type (e.g. $MoS_2$) or p-type (e.g. $WSe_2$) SC-TMDs available or the type of charge carriers can be induced by contacting the materials by metals with different work functions [160]. In such a way, atomically thin p-n diodes [57] or electrically tunable photovoltaic devices [60] can for instance be realized.

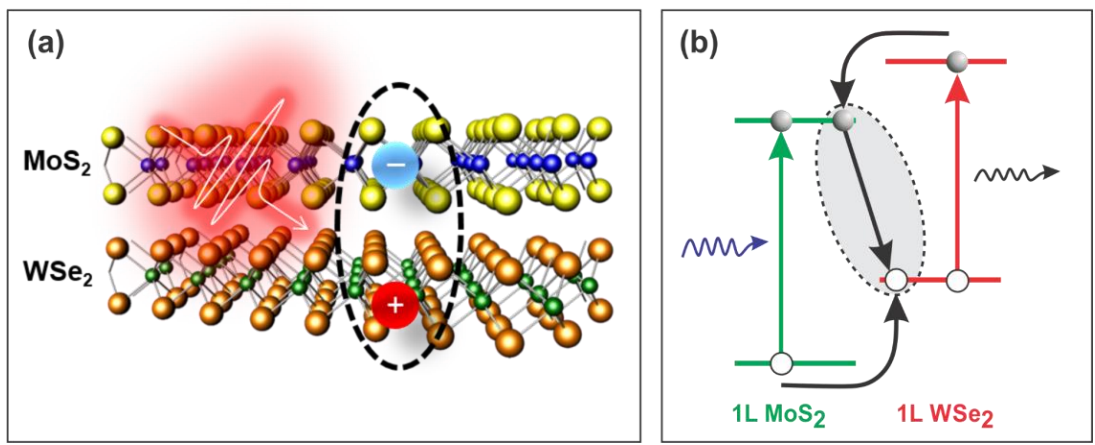

*Figure 9: (a) Cartoon depicting a vertical van der Waal heterostructure consisting of monolayers of MoS$_2$ and WSe$_2$ with a photo-excited interlayer exciton (IX) with the electron localized in MoS$_2$ and the hole localized in the WSe$_2$ layer. (b) Scheme of the type-II band alignment of valence and conduction bands at the K- and K'-points of stacked MoS$_2$ and WSe$_2$ monolayers. The process of photo-excitation of charge carriers, charge carrier relaxation across the heterostructure, formation of bound*

Two-dimensional materials offer a major advantage compared to three-dimensional solid state based heterostructures such as GaAs/AlGaAs: due to the absence of dangling bonds on the surface, 2D materials can be arbitrarily combined with atomistically precise interfaces and with an additional rotational degree of freedom [11,104,161]. Such 2D heterostructures can be prepared either by van der Waals assembly of micromechanically exfoliated or grown 2D crystals [13,99] [sketched in Fig. 9(a)] or by a direct growth of lateral and vertical 2D heterostructures [119,120]. The precise control of the interface is a challenge that needs to be overcome in order to build functional heterostructures with a sufficient interlayer coupling. Hereby, not only the rotational control is of importance, but also strategies need to be developed to avoid and remove contamination from the interfaces. The interface can either be polluted by thin water films or by polymer residues from the carrier materials used during the transfer process. In order to maintain a sufficient interlayer coupling, careful cleaning of the interface with suitable solvents, annealing procedures or performing the transfer process in inert atmosphere are possible routes. Interlayer coupling in vertical van der Waals heterostructures has been investigated by means of optical spectroscopy with reflectance [162], photo-luminescence [161,163–165], electroluminescence [58], and Raman-experiments[155], by measurements of the photovoltaic effect [57,60] or by scanning tunneling spectroscopy [166,167].

Fig. 9(b) schematically depicts a vertical van der Waals heterostructure built from $MoS_2$ and $WSe_2$ monolayers with type-II band alignment. The figure also sketches the process for the separation of photo-excited charge carriers across a heterojunction, the formation of interlayer excitons and the radiative re-emission. Incoming photons are absorbed in both monolayers, and electrons are excited from the valence band to the conduction band. At the heterostructure-interface the electrons are transferred to the favorable lower potential conduction band in $MoS_2$, while the holes are transferred to the favorable higher potential valence band in $WSe_2$. The charge dynamics cause a separation of the *e-h* pairs into different layers. If the Coulomb-interaction between the electron and holes in the different layers is strong enough, bound bosonic quasiparticles called interlayer excitons (IX) can form. The recombination of such an IX occurs under emission of a photon with characteristic energy that is different to the energy of an exciton stemming from the two individual layers. The band alignment of conduction band minima and valence band maxima vs. the vacuum level at the *K*- and *K'*- points of the direct gap monolayers of $MoS_2$, $WSe_2$, $MoSe_2$, and $WS_2$ are summarized in Fig. 10. It is obvious that for any possible combination of two different monolayers, a type-II band heterojunction forms with the above described mechanism for e-h separation, whereas the electrons and holes can be either electrically injected or photo-excited. By application of an in-plane electric field, the charge carriers can either be forced away towards the contacts as required for photovoltaic applications [60] or pushed towards the heterostructures' interface as it is needed for light-emitting diodes.

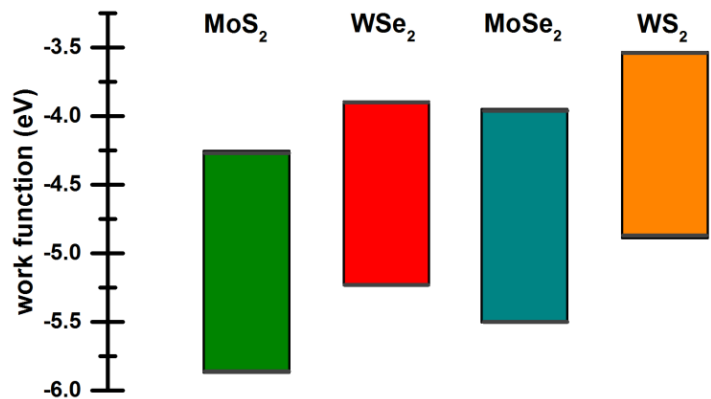

Figure 10: Work function and band alignement of valence and conduction band edges at the K- and K'- points for $MoS_2$, $WSe_2$, $MoSe_2$, and $WS_2$ mono-layers (values taken from [162]). All possible combinations of vertically stacked van-der Waals hetero-bilayers result in a typ-II band alignement.

Fig. 11(a) shows an optical micrograph of a typical van der Waals heterostructure stacked from micromechanically exfoliated $MoS_2$ and $WSe_2$ layers, which were fabricated by the so-called dry transfer method using a PDMS stamp [101]. It has been proven by Raman investigations [Fig. 11 (b)] that the several µm large area of the heterostructure is prepared from monolayers. Fig 11 (c) depict the emission spectra from a $MoS_2$ mono-layer, a $WSe_2$ monolayer and the heterostructure region at room temperature. The weaker emission from $MoS_2$ occurs at an energy of 1.81 eV. The emission of $WSe_2$ at 1.64 eV exhibits a red-shifted shoulder that is assigned to the emission of a charged exciton (trion) [24]. The emission from the heterostructure is red-shifted with a peak energy of 1.58 eV in agreement with earlier reports [161,163]. We would like to emphasize that the red-shifted emission is constant over the whole heterostructure area on the sample. Compared to the emission from $WSe_2$, the intensity from the heterostructure is reduced by a factor of six. Particularly, the reduced emission energy is a strong hint that the coupling between the two TMD monolayers is strong enough to form a type-II heterostructure causing an effective e-h separation and the formation of interlayer excitons even at room temperature. Hereby, TMD-heterostructures represent an intriguing possibility to excitonic phenomena compared to conventional heterostructures e.g. in GaAs with a much smaller exciton binding energy of only a few meV [168].

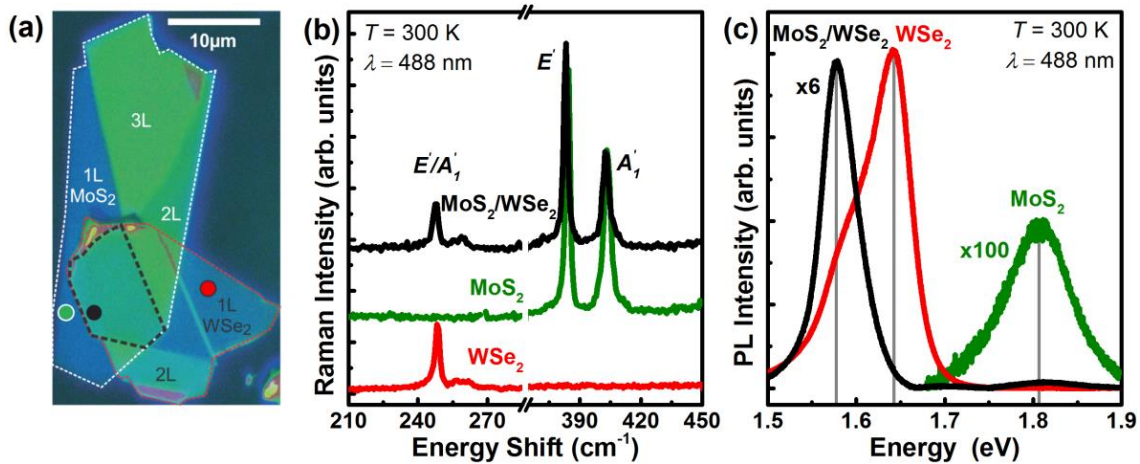

*Figure 11: (a) False color representation of an optical micrograph of a vertically stacked $MoS_2$/$WSe_2$ heterostructure prepared by a dry transfer of micromechanically exfoliated flakes. The region of the hetero-bilayer area consisting of a monolayer $MoS_2$ and a monolayer $WSe_2$ is marked by a blacked dashed line. The positions for optical measurements are indicated with black (hetero-bilayer) green ($MoS_2$ monolayer) and red ($WSe_2$ monolayer). (b) Raman spectra from $MoS_2$/$WSe_2$ heterostructure (black), $MoS_2$ monolayer (green) and $WSe_2$ monolayer displaying the Raman active high frequency E' and A'$_1$ phonon modes with the suitable energies for the related monolayers (excited at room temperatures in vacuum with $\lambda_{Laser}$ = 488 nm). (c) Photoluminescence (PL) spectra from $MoS_2$/$WSe_2$ heterostructure (black), $MoS_2$ monolayer (green) and $WSe_2$. The red-shifted photoluminescence peak from the heterostructure is a strong hint for the formation of interlayer excitons. (Raman and PL spectra taken at room temperatures in vacuum with $\lambda_{Laser}$ = 488 nm).*

The IX excitons in vertical van der Waals heterostructures are long-lived with a lifetime of up to $\tau$ = 1.8 ns (for MoSe$_2$-WSe$_2$ heterostructures at $T$ = 20 K) [164]. This lifetime is orders of magnitude longer than the lifetime of the direct excitons in the individual SC-TMDs monolayers that are in the range of tens of ps [169,170]. The increased lifetime can be explained by a reduced spatial overlap of the wave functions from electron and holes residing in different layers. Therefore, the interlayer excitons are of spatially indirect nature with reduced optical dipole moments similar to those created e.g. in conventional GaAs based double quantum well structures with a perpendicular applied electric field [171]. It has been shown that also for van der Waals heterostructures the IX emission energy can be controlled by the application of an electric field using top and bottom gate electrodes [164].

Long-lived excitons are of great interest to study correlation and quantum phenomena in dense ensembles of bosonic quasiparticles [61,172]. Such a multi-exciton system exhibits a rich phase diagram depending on temperature and quasiparticle density with different phases from a classical exciton gas to a bosonic exciton gas that can condense to a Bose-Einstein condensate (BEC) by further cooling [61,172]. A BEC is a coherent state of bosons in the same ground state. BEC of excitons in solid state systems is a subject of intense research thus far concentrated on low-dimensional heterostructures embedded in conventional three-dimensional crystals studied below 4K [172–175]. It has been predicted by numerical investigations together with scaling arguments that hetero-bilayers prepared from SC-TMDs with an insulating interlayer prepared from atomically thin hexagonal boron nitride sheets (hBN) could result in a degenerate Bose gas with a superfluid phase with record high temperature approaching 100 K [61].

This underlines that SC-TMD heterostructures are not only highly interesting and of significant potential from the application point of view but also to study correlation and many-body quantum phenomena in two-dimensional systems.

## *6. Conclusion and perspectives*

Atomically thin crystalline membranes are promised as platform for novel technologies and also fundamental studies. The materials with manifold electronic, optical, mechanical and thermal properties are inherently fascinating due to coincidence of the dimensionality in real and momentum space – they are truly two-dimensional membranes. The absence of dangling bonds at the interface, 2D materials can be combined to vertical van der Waals heterostructures and integrated in 3D hybrid networks without limitation. The direct access of the 2D surface enables interfacial engineering and environmental sensing as well as a high degree of tunability and a good electrostatic control etc. The integration of 2D materials is not restricted to solid state materials due to the flexibility of most 2D materials. The material in the focus of this review are SC-TMDs with a direct band gap in the monolayer limit in the visible range. The materials feature high sunlight absorption up to 15% for a single monolayer, high stability, catalytic activity, and a decent charge carrier mobility. They are stretchable and bendable. Most of fundamental studies are still carried out on micromechanical exfoliated individual flakes. However, material scientist are progressing quickly in developing routes for large scale production. In addition, the SC-TMDs provide access to a variety of structural and electronic properties such as defects, strain, number of layers, doping and temperature by studying the higher frequency Raman active phonon modes. The impressive scattering cross-section for only single crystal layer makes inelastic light scattering experiments a versatile, non-destructive, contactless tool for advanced characterization without the need to prepare electric contacts or similar. Also the excitonic properties of the materials are superior. Stacked as van der Waals heterostructures, a heterojunction with type-II band alignment forms required for several applications including photovoltaics, effective photo-catalytics, light-emitting diodes, but also suitable to generate p-n junctions. Moreover, long-lived interlayer excitons can be photoexcited with the charge carriers separated in the different layers. A reduced spatial overlap of the wave functions increases the IX lifetime and enables the investigation of correlated phenomena and many-body quantum effects in dense exciton ensemble. A peculiar example is the condensation of the bosonic quasiparticle to a coherent phase that is predicted to happen at record high temperatures in SC-TMD heterostructures [164].

With all these 'superlatives' it remains thrilling to learn about the first real device made out of SC-TMD on the market that can be bought - different as the application as dry lubricant. Such a device must either be much cheaper in the production as already exciting products on the marked or device characteristics must exceed those of existing devises by a lot or a completely new functionality must be achievable with the SC-TMDs. For these reasons, we surmise that it will not be in electronic devices, where SC-TMD based field effect transistor have comparable key characteristics compared to silicon-based devices, but the silicon technology is highly industrialized. On the basis of the above given considerations, the first devices might be in the area of (photo-activated) bio-sensors, gas-sensor, bio-medical application, solar energy conversation or photo-catalysis. But also on the fundamental side SC-TMDs hold numerous exciting properties that are worth to be explored further such as doping induced superconductivity in less than 1 nm thin crystals, many-body interactions in exciton ensembles, spin- and valley-properties and their control and interaction e.g. in hybrid devices, defect engineering and the controlling of single photon emitters, defect engineering, integration of SC-TMDs with strong light matter interaction in photonic and plasmonic circuitry, just to mention some possible directions. There will be an exciting future in a highly active and interdisciplinary field by studying optically active SC-TMDs and other 2D materials.

*Acknowledgements*

We acknowledge financial support by Deutsche Forschungsgemeinschaft (DFG) via excellence cluster 'Nanosystems Initiative Munich' (NIM), through the TUM International Graduate School of Science and Engineering (IGSSE) and DFG projects WU 637/4-1, HO3324/9-1 and BaCaTeC.